\documentclass[twocolumn,showpacs,amssymb]{revtex4}
\usepackage{natbib}
\usepackage{epsfig}
\usepackage{graphicx}
\usepackage{float}
\usepackage{amsmath}
\usepackage{color}
\usepackage{amssymb}
\usepackage{amsfonts}
\usepackage{units}
\usepackage{dcolumn}
\usepackage{bm}
\usepackage[colorlinks,linkcolor=blue,anchorcolor=green,citecolor=blue]{hyperref}

\begin{document}

\title{New test of weak equivalence principle using polarized light from astrophysical events}

\author{Xue-Feng Wu$^{1,2,3\ast}$, Jun-Jie Wei$^{1,4}$, Mi-Xiang Lan$^{1}$, He Gao$^{5}$, Zi-Gao Dai$^{6}$, and Peter M{\'e}sz{\'a}ros$^{7,8,9}$}

\affiliation{$^1$ Purple Mountain Observatory, Chinese Academy of Sciences, Nanjing 210008, China \\
$^2$ School of Astronomy and Space Science, University of Science and Technology of China, Hefei, Anhui 230026, China \\
$^3$ Joint Center for Particle, Nuclear Physics and Cosmology, Nanjing University-Purple Mountain Observatory, Nanjing 210008, China\\
$^4$ Guangxi Key Laboratory for Relativistic Astrophysics, Nanning 530004, China \\
$^5$ Department of Astronomy, Beijing Normal University, Beijing 100875, China\\
$^6$ School of Astronomy and Space Science, Nanjing University, Nanjing 210093, China\\
$^7$ Department of Astronomy and Astrophysics, Pennsylvania State University, 525 Davey Laboratory, University Park, Pennsylvania 16802, USA\\
$^8$ Department of Physics, Pennsylvania State University, 104 Davey Laboratory, University Park, Pennsylvania 16802, USA\\
$^9$ Center for Particle and Gravitational Astrophysics, Institute for Gravitation and the Cosmos, Pennsylvania State University, 525 Davey
Laboratory, University Park, Pennsylvania 16802, USA\\
$^\ast$Electronic address: xfwu@pmo.ac.cn}

\date{\today}


\begin{abstract}
Einstein's weak equivalence principle (WEP) states that any freely falling, uncharged test particle
follows the same identical trajectory independent of its internal structure and composition. Since
the polarization of a photon is considered to be part of its internal structure, we propose that
polarized photons from astrophysical transients, such as gamma-ray bursts (GRBs) and fast radio
bursts (FRBs), can be used to constrain the accuracy of the WEP through the Shapiro time delay effect.
Assuming that the arrival time delays of photons with different polarizations are mainly attributed
to the gravitational potential of the Laniakea supercluster of galaxies, we show that a strict upper
limit on the differences of the parametrized post-Newtonian parameter $\gamma$ value
for the polarized optical emission of GRB 120308A is $\Delta\gamma<1.2\times10^{-10}$,
for the polarized gamma-ray emission of GRB 100826A is $\Delta\gamma<1.2\times10^{-10}$,
and for the polarized radio emission of FRB 150807 is $\Delta\gamma<2.2\times10^{-16}$.
These are the first direct verifications of the WEP for multiband photons with different polarizations.
In particular, the result from FRB 150807 provides the most stringent limit to date on a deviation
from the WEP, improving by one order of magnitude the previous best result based on Crab pulsar
photons with different energies.
\end{abstract}

\maketitle

\section{Introduction}

Einstein's weak equivalence principle (WEP) is a fundamental postulate of general relativity and of other
matric theories of gravity. One statement of the WEP is that the trajectory of any freely falling, uncharged
test body is independent of its internal structure and composition \cite{wil06}. In the simplest case of
considering two different bodies in a gravitational field, the WEP states that these two bodies fall with
the same acceleration. Put differently, any two different kinds of massless (or negligible rest mass) neutral
particles, or two of the same particles with different internal structures (e.g., energies or polarizations)
or different compositions, if emitted simultaneously from the same astronomical source and traveling across
the same gravitational field, should be received at the same time by the observer.
The energy and polarization independence of photons and electromagnetic (EM) wave packet
propagation in spacetime is one of the consequences of the WEP.

In the parametrized post-Newtonian (PPN) formalism, the validity of the WEP can be characterized by
constraints on the differences in the PPN parameters (e.g., the parameter $\gamma$ which represents the
level of space curvature per unit rest mass) for different particles, since all gravity theories satisfying
the WEP predict $\gamma_{1}=\gamma_{2}\equiv\gamma$, where the subscripts correspond to two different
particles \cite{wil06}. According to the Shapiro (gravitational) time delay effect \cite{sha64}, the time
interval for particles to pass through a given distance is longer by
\begin{equation}
t_{\rm gra}=-\frac{1+\gamma}{c^3}\int_{r_0}^{r_e}~U(r)dr
\end{equation}
in the presence of a gravitational potential $U(r)$, where $r_e$ and $r_o$ denote the locations of the source
and the observer, respectively. If the WEP fails, the $\gamma$ values will no longer be the same for different particles,
leading to the two particles emitted simultaneously from the same source to have different arrival times.
The relative Shapiro time delay is then given by
\begin{equation}
\Delta t_{\rm gra}=\frac{\gamma_{\rm 1}-\gamma_{\rm 2}}{c^3}\int_{r_o}^{r_e}~U(r)dr\;,
\label{gra}
\end{equation}
where the difference of the $\gamma$ values $\Delta\gamma=\gamma_{\rm 1}-\gamma_{\rm 2}$ can be used as a
measure of a possible violation of the WEP.

Up to now, the observed time delays of different types of messenger particles (e.g. photons, neutrinos, or
gravitational waves), or of the same types of particles but with different energies from astronomical sources
have been used to test the accuracy of the WEP through the relative differential variations of the $\gamma$
values, such as the particle emissions from supernovae 1987A \cite{Krauss88,Longo88}, gamma-ray bursts (GRBs)
\cite{gao15,wei2016,Sang2016}, fast radio bursts (FRBs) \cite{wei15,tin16}, blazars \cite{wei16,Wang2016},
the Crab pulsar \cite{yang16,zhanggong16}, and gravitational wave (GW) sources \cite{wu2016,Kahya2016}.
Particularly, with the assumption that the arrival time delays of
FRB photons with different energies are caused dominantly by the gravitational potential of the Milky Way,
Ref. \cite{wei15} placed a stringent limit on $\gamma$ differences of $\Delta \gamma<4.36\times10^{-9}$,
improving the previous results by 1--2 orders of magnitude.
Moreover, it has been shown that much more severe constraints on the WEP accuracy can be achieved
($\Delta \gamma \sim 10^{-13}$) when considering the gravitational potential of the large-scale structure
rather than the gravity of the Milky Way \cite{nus16,Zhang2016}.
Most recently, Ref. \cite{yang16} showed that a giant pulse from the Crab pulsar with a 0.4-ns residual time delay
between energies set the current best limit on a deviation from the WEP of
 $\Delta\gamma < (0.6-1.8)\times 10^{-15}$, even though the Crab pulsar is in the Milky Way.

The photon is generally viewed as a massless bundle of EM
energy and the photon momentum is defined as the ratio between the photon energy and the speed of light.
Maxwell's theory implies that EM radiation carries both energy and momentum \cite{Jackson1978}.
The Poynting vector \textbf{E$\times$B} gives the linear momentum density, and the cross product
of the Poynting vector with the position vector represents the angular momentum. Furthermore, circularly
polarized light can have angular momentum, which is defined as the ratio between the free energy
per unit volume and the angular frequency. Thus, the photon angular momentum has two components:
one is the spin angular momentum, depending on the polarization; the other is the orbital angular momentum,
which is independent of the polarization but depends on the spatial distribution \cite{Jackson1978}.
Regarding the nature or the internal structure of the photon, that has been discussed in detail by
Ref. \cite{Roych2008}.
The main views concerning an understanding of the photon structure are as follows. Einstein considered the
photon as a singular point which is surrounded by EM fields. In quantum electrodynamics, the photon is viewed
as a unit of excitation associated with the quantized mode of the radiation field, characterized by a precise
momentum, energy, and polarization. In sum, the polarization can be considered as one of the parameters
characterizing the internal structure of photons (see, e.g., Ref. \cite{Hofer1998}).
Furthermore, the polarization of photons is a common and interesting observational feature of some transient events.
Thus, measurements of polarization can not only be used to explore the nature of the transient, but may also
be used to test the accuracy of the WEP (more on this below).
It has been proposed that polarization is important in verifying the Einstein equivalence principle
(see, e.g., Ref. \cite{di Serego Alighieri2015}). It is worth pointing out that some measurements of polarized light
are, in fact, polarized EM waves, i.e., a collective phenomenon as opposed to a property of the individual photon.
For such a case, we suggest two EM waves with different polarizations can also be used to test the WEP.
Similarly, besides two correlated photons with different energies, two light curves in different energy bands have
also been used to test the WEP.

Tests of the WEP at the post-Newtonian level have reached a high precision, but we note that all the tests so far
have relied on the relative arrival time delays of the same species of particle with varying energies or
of different species of messenger particles. Since the WEP emphasizes that any freely falling, uncharged test
body will follow a trajectory independent of its internal structure and composition \cite{wil06},
multiband EM emissions exploiting different internal structures (e.g, polarizations) are an essential
tool for further testing the WEP to a higher accuracy level.
Here we propose for the first time that the time delays of photons and EM waves with different polarizations from
astrophysical transients, such as GRBs and FRBs, can also be used to constrain the WEP, thus extending the
tested WEP internal structure to the polarization and thereby obtaining the strictest limit so far on
the WEP.

\section{Method description}

To calculate the relative Shapiro time delay with Eq.~(\ref{gra}), one needs to know the gravitational potential
$U(r)$ along the propagation path.  Recent studies suggest that incorporating the large-scale gravitational
potential can provide better constraints on the WEP accuracy \cite{nus16}.
Laniakea is a newly discovered supercluster of galaxies, of which our Milky Way galaxy as well as the Local Group
are part  \cite{{Tully2014}}.
Laniakea, if approximated as round, has a diameter of $\sim160$ Mpc, and encloses $\sim10^{17}$ solar masses.
As long as the distance of an astrophysical event is far beyond the scale of Laniakea, it is reasonable to
adopt in our calculations the gravitational potential of the Laniakea supercluster of galaxies as being dominant
potential. Strictly speaking, $U(r)$ has contributions from the gravitational potentials of the transient host
galaxy $U_{\rm host}(r)$, the intergalactic background $U_{\rm IG}(r)$, and the Laniakea supercluster of galaxies
$U_{\rm L}(r)$. Since the potential models of $U_{\rm host}(r)$ and $U_{\rm IG}(r)$ are extremely uncertain,
and the contribution from $U_{\rm L}(r)$ can dominate the other two components, we consider only the
potential of the Laniakea $U_{\rm L}(r)$.

Assuming that the observed time delay between two polarized photons from the same astrophysical source
is mainly attributed to the gravitational potential of the Laniakea supercluster of galaxies,
and adopting a Keplerian potential $U_{\rm L}(r)=-GM/r$ for Laniakea \cite{footnote}, one therefore has
(see Refs. \cite{Longo88,wei2016} for more details)
\begin{equation}\label{gammadiff}
\begin{aligned}
\Delta t_{\rm obs} &>\Delta t_{\rm gra}= \Delta\gamma \frac{GM_{\rm L}}{c^{3}} \times \\
          &\ln \left\{ \frac{ \left[d+\left(d^{2}-b^{2}\right)^{1/2}\right] \left[r_{L}+s_{\rm n}\left(r_{L}^{2}-b^{2}\right)^{1/2}\right] }{b^{2}} \right\}\;,
\end{aligned}
\end{equation}
where $M_{\rm L}\simeq10^{17}M_{\odot}$ is the total mass of Laniakea \cite{Tully2014},
$b$ corresponds to the impact parameter of the light path relative to the Laniakea center,
$d$ is the approximate distance from the transient to Earth,
$r_{L}\simeq77$ Mpc denotes the distance from the Laniakea center to Earth \cite{Lynden-Bell1988},
and $s_{\rm n}$ is the sign of the correction of the source direction, where
$s_{\rm n}=+1$ ($s_{\rm n}=-1$) stands for the source located along the direction of Laniakea (anti-Laniakea) center.
\emph{Note that the total mass of Laniakea ($M_{\rm L}\simeq10^{17}M_{\odot}$) includes
the contributions from dark matter as well as baryonic matter. Only a minor fraction of the
matter is baryonic \cite{Tully2014}.}

\section{WEP tests with polarized light}

With Eq.~(\ref{gammadiff}), one can test the WEP by setting a strict limit on $\Delta\gamma$.
As can be seen from Eq.~(\ref{gammadiff}), the shorter time delay between two polarized photons or EM waves
and the larger the distance of the transient, the better the constraint on the WEP.
GRBs and FRBs are two common transients where  measurements of the polarization are available.
In this work, we use the time delays between different polarized light from GRBs and FRBs to constrain the WEP.

\subsection{Polarized optical or gamma-ray emission from GRBs}

GRBs are the most energetic explosions occurring at cosmological distances. The polarimetric observations
are particularly important, because they provide us with completely different information about the GRB jets
and central engines. The first polarization measurements of GRBs were performed in the optical afterglows.
There have been only a few optical afterglows with polarized emission measured in the past decade,
including the following representative cases. Reference \cite{Greiner2003} reported the temporal evolution
of the polarization degree and polarization angle for the optical afterglow of GRB 030329.
Reference \cite{Mundell2007} set an upper limit on the polarization degree ($\Pi<8\%$) for the early optical afterglow
of GRB 060418. Reference \cite{Steele2009} reported a polarization degree of $\Pi=10\pm1\%$ for the early optical afterglow
of GRB 090102. Reference \cite{Mundell2013} detected an evolving polarization degree and a nearly constant polarization
position angle in the optical band of GRB 120308A.

On the other hand, there have been also some polarization detections in the prompt gamma-ray emission of GRBs.
The first report was the measurement of a high linear polarization degree of $\Pi=80\pm20\%$ from GRB 021226
\cite{Coburn2003}. However, a subsequent reanalysis of the same data could not confirm
any polarization signal \cite{Rutledge2004}.
A second report, on GRB 041219A, also indicated a high polarization degree, with
$\Pi=98\pm33\%$ \cite{Kalemci2007} and $\Pi=68^{+31}_{-30}\%$ \cite{McGlynn2007}.
Using a different instrument, the Gamma-Ray Burst Polarimeter (GAP) on board the Interplanetary Kite-craft
Accelerated by Radiation of the Sun (IKAROS), Reference \cite{Yonetoku2011} reported a polarization detection in
the prompt gamma-ray emission of GRB 100826A. They detected a significant change of polarization angle
with a $3.5\sigma$ confidence level and the average polarization degree of $\Pi=27\pm11\%$ with a $2.9\sigma$
confidence level.  Two other highly polarized bursts, GRB 110301A and GRB 110721A, with polarization degrees
of $\Pi=70\pm22\%$ ($3.7\sigma$) and $\Pi=84^{+16}_{-28}\%$ ($3.3\sigma$), respectively, were also detected
by the GAP instrument \cite{Yonetoku2012}.

As examples, we use here two polarization measurements of GRBs to constrain the WEP, namely, the
polarized optical emission from GRB 120308A and the polarized gamma-ray emission from GRB 100826A.

GRB 120308A was detected by the Swift satellite on 2012 March 8 at $T_{0}=06:13:38$ UT, with coordinates (J2000)
R.A.=$14^{\rm h}36^{\rm m}20^{\rm s}.38$ and Dec.=$+79^{\circ}41^{'}10^{''}.6$ \cite{Baumgartner2012}. Its redshift
is $z=2.2$ \cite{Mundell2013}. Reference \cite{Mundell2013} reported the detection of a polarization degree of
$\Pi=28^{+4}_{-4}\%$ in the immediate optical afterglow of GRB 120308A, 4 min after its trigger in the
gamma-ray band, decreasing to $\Pi=16^{+5}_{-4}\%$ over the subsequent 10 min. The polarization angle
remained stable, changing by no more than $15^{\circ}$ over this time. The arrival lag $\Delta t_{\rm obs}$
for optical photons ranging in polarization degree from $\Pi=28^{+4}_{-4}\%$ to $\Pi=16^{+5}_{-4}\%$ is 6 min.
With the above information on GRB 120308A, we thus obtain a WEP constraint from Eq.~(\ref{gammadiff})
\begin{equation}
\Delta\gamma<1.2\times10^{-10}\;.
\end{equation}

The GAP on board IKAROS was designed to detect the degree of linear polarization in the prompt emission of
GRBs in the energy range of 70--300 keV. GRB 100826A was detected by the GAP on 2010 August 26 at
$T_{0}=22:57:20.8$ UT, with coordinates (J2000) R.A.=$279.6^{\circ}$ and Dec.=$-22.3^{\circ}$ \cite{Yonetoku2011}.
Since an optical counterpart of this GRB was not identified, its redshift is unknown. Following the treatment of
Ref. \cite{Deng2014}, we use the luminosity relation to estimate its redshift and obtain $z\geq0.054$.
To account for the uncertainty of the redshift estimates, Ref. \cite{wei15} tested the results
by varying the source distance from 1 Mpc (the distance to the edge of the Local Group) to 3$z_{\rm infer}$.
As shown in Fig.~1 of Ref. \cite{wei15}, they found that even if the distance estimates for the sources
have large uncertainties, the results on testing the WEP will not be significantly affected,
i.e., the constraint results vary within one order of magnitude or less.
Reference \cite{Yonetoku2011} divided the light curve of the prompt emission of GRB 100826A into two time intervals
(denoted by interval-1 and -2) for the polarization analysis.
The first interval of this burst shows a large flare lasting 47 s since the trigger time, and the second one consists
of several spikes lasting 53 s. The best values of the polarization degrees and the polarization angles
($\phi$) are $\Pi_{1}=25\pm15\%$ with $\phi_{1}=159\pm18^{\circ}$ for interval-1 and $\Pi_{2}=31\pm21\%$ with
$\phi_{2}=75\pm20^{\circ}$ for interval-2, respectively. That is, the average polarization degree over the burst
duration is $\Pi=27\pm11\%$ in the energy range of 70--300 keV,
and the polarization angle significantly varies from interval-1 to -2.
To be conservative, we adopt the sum of the two time intervals, 100 s, as the arrival lag $\Delta t_{\rm obs}$
for gamma-ray photons ranging in polarization angle from $\phi_{1}=159\pm18$ to $\phi_{2}=75\pm20^{\circ}$.
With the inferred redshift of $z=0.054$, a strong limit on the WEP from
Eq.~(\ref{gammadiff}) is
\begin{equation}
\Delta\gamma<1.2\times10^{-10}
\end{equation}
for GRB 100826A.

\subsection{Polarized radio emission from FRBs}

As a new and highly unusual type of millisecond radio transients, FRBs are one of the most discussed astronomical
phenomena of recent years.  So far, only 18 FRBs have been reported \cite{Lorimer2007,Keane2016,Petroff2015,Masui2015,Ravi2016},
and their observed event rate is estimated to be $\sim10^{-3}~\unit{galaxy^{-1}}~\unit{yr^{-1}}$. Most of them are
characterized by high galactic latitudes and large dispersion measures (DMs), which strongly suggest that FRBs are
of an extragalactic or even cosmological origin.
The first claim to measure a redshift was made for FRB 150418 based on the identification of a fading
radio transient \cite{Keane2016}. However, the redshift measurement of FRB 150418 has subsequently been challenged \cite{Williams2016}.
Very recently, the observation of repeated emissions from FRB 121102 \cite{Chatterjee2017} have permitted the precise localization
of its host galaxy, which has made a precise redshift determination to FRB 121102, $z=0.19273(8)$ \cite{Tendulkar2017}.
The polarimetric observations show that photons in the radio emission of some FRBs are polarized. For instance,
the first reported polarization measurement was from FRB 140514, which was found to be $21\pm7\%(3\sigma)$ circularly
polarized on the leading edge with a $1\sigma$ upper limit ($<10\%$) on the linear polarization \cite{Petroff2015}.
A strong linear polarization was detected in FRB 110523, with a linear polarization fraction of $44\pm3\%$,
and its Faraday rotation measure was then determined by the linear polarization \cite{Masui2015}.
The highest linearly polarized burst reported to date was FRB 150807, which had an $80\pm1\%$ linear polarization
fraction \cite{Ravi2016}.
Here we take the highest linearly polarized burst (FRB 150807) as an example and use its polarization
information to constrain the WEP.

FRB 150807 was detected by the 64-m Parkes radio telescope
on 2015 August 7 at 17:53:55.78 UT over the 1182--1519.5 MHz band, with coordinates (J2000)
R.A.=$22^{\rm h}40^{\rm m}23^{\rm s}$ and Dec.=$-53^{\circ}16^{'}$ \cite{Ravi2016}.
The line-of-sight free electron column density of FRB 150807, measured in units of the DM,
is $266.5\pm0.1~\unit{cm^{-3}~pc}$, which is one of the smallest values reported for a FRB.
However, this still greatly exceeds the expected foreground Milky Way DM, estimated to be $70\pm20~\unit{cm^{-3}~pc}$.
After removing the Milky Way DM contribution, the extragalactic DM is about $196.5~\unit{cm^{-3}~pc}$.
With this extragalactic DM value, FRB 150807 is inferred to be at a redshift $z=0.16$
(corresponding to a comoving distance of $d=660$ Mpc) \cite{web}. Reference~\cite{Ravi2016} estimated the distance of
FRB 150807 through its localization. The deepest archival images of the sky-localization area include
nine bright objects: three stars and six galaxies. If FRB 150807 originated in a galaxy, it is expected to be
$d>500$ Mpc distant \cite{Cordes2016}, in good agreement with the value inferred from the extragalactic DM.
The lower limit on the  distance ($d=500$ Mpc) is therefore conservatively adopted for the rest of this paper.
From the polarization evolution of FRB 150807 (see Fig.~1 of Ref.~\cite{Ravi2016}), one can easily identify the
arrival time delay $\Delta t_{\rm obs}\simeq0.256$ ms for radio photons ranging in polarization angle from
about $-31.2^{+0.82}_{-0.77}$ to $-34.0^{+0.82}_{-0.86}$ deg. We thus obtain the WEP constraint from
Eq.~(\ref{gammadiff}) for FRB 150807
\begin{equation}
\Delta\gamma<2.2\times10^{-16}\;,
\end{equation}
which is almost 10 times tighter than the previous best limit from the Crab pulsar value derived using photons
with different energies, which was $\Delta\gamma \sim 10^{-15}$ \cite{yang16}. All above-mentioned limits on
$\Delta\gamma$ through the Shapiro time delay effect are listed in Table~\ref{table1} for comparison.

\begin{table*}[h]
{\tiny
\caption{Upper bounds on the differences of the $\gamma$ values from the Shapiro time delay measurements.}
\begin{tabular}{lllccc}
&&&&& \\
\hline\hline
&&&&& \\
{Author (year)}&{Source}&{Messengers}&{Gravitational field}&{$\Delta\gamma$}&{References}\\
&&&&& \\
\hline
Krauss \& Tremaine (1988) & Supernova 1987A & eV photons and MeV neutrinos & Milky Way & $5.0\times10^{-3}$ & \cite{Krauss88} \\
Longo (1988) & Supernova 1987A & eV photons and MeV neutrinos & Milky Way & $3.4\times10^{-3}$ & \cite{Longo88} \\
             & Supernova 1987A & 7.5--40 MeV neutrinos & Milky Way & $1.6\times10^{-6}$ & \cite{Longo88} \\
Gao \emph{et al.} (2015)  & GRB 090510 & MeV--GeV photons & Milky Way & $2.0\times10^{-8}$ & \cite{gao15} \\
                          & GRB 080319B & eV--MeV photons & Milky Way & $1.2\times10^{-7}$ & \cite{gao15} \\
Wei \emph{et al.} (2015)  & FRB 110220 & 1.2--1.5 GHz photons & Milky Way & $2.5\times10^{-8}$ & \cite{wei15} \\
                          & FRB/GRB 100704A & 1.23--1.45 GHz photons & Milky Way & $4.4\times10^{-9}$ & \cite{wei15} \\
Tingay \& Kaplan (2016)   & FRB 150418 & 1.2--1.5 GHz photons & Milky Way & (1--2)$\times10^{-9}$ & \cite{tin16} \\
Nusser (2016)             & FRB 150418 & 1.2--1.5 GHz photons & Large-scale structure & $10^{-12}$--$10^{-13}$ & \cite{nus16} \\
Wei \emph{et al.} (2016a)  & Blazar Mrk 421 & keV--TeV photons & Milky Way & $3.9\times10^{-3}$ & \cite{wei16} \\
                           & Blazar PKS 2155-304 & sub TeV--TeV photons & Milky Way & $2.2\times10^{-6}$ & \cite{wei16} \\
Wang \emph{et al.} (2016)  & Blazar PKS B1424-418 & MeV photons and PeV neutrino & Virgo Cluster & $3.4\times10^{-4}$ & \cite{Wang2016} \\
                           & Blazar PKS B1424-418 & MeV photons and PeV neutrino & Great Attractor & $7.0\times10^{-6}$ & \cite{Wang2016} \\
Wei \emph{et al.} (2016b)  & GRB 110521B & keV photons and TeV neutrino  & Laniakea supercluster of galaxies & $1.3\times10^{-13}$ & \cite{wei2016} \\
Wu \emph{et al.} (2016a)  & GW 150914 & 35--150 Hz GW signals  & Milky Way  & $\sim10^{-9}$ & \cite{wu2016} \\
Yang \& Zhang (2016)  & Crab pulsar & 8.15--10.35 GHz photons  & Milky Way  & (0.6--1.8)$\times10^{-15}$ & \cite{yang16} \\
Wu \emph{et al.} (2016b)  & GRB 120308A & Polarized optical photons  & Laniakea supercluster of galaxies  & $1.2\times10^{-10}$ & This paper \\
                          & GRB 100826A & Polarized gamma-ray photons  & Laniakea supercluster of galaxies  & $1.2\times10^{-10}$ & This paper \\
                          & FRB 150807 & Polarized radio photons  & Laniakea supercluster of galaxies  & $2.2\times10^{-16}$ & This paper \\
\hline\hline
\end{tabular}
\label{table1}
}
\end{table*}

\section{Conclusions}

The fact that the trajectory of any freely falling, uncharged test body is independent of its internal structure
is one of the consequences of the WEP. Since the polarization is considered a basic component of the internal structure
of photons, we have proposed that polarization measurements of astrophysical transients, such as GRBs and FRBs, can
provide stringent tests of the accuracy of the WEP.
In other words, the validity of the WEP can be tested with the arrival time delays between photons with different
polarizations.  With the assumption that the time delays are solely caused by the gravitational potential of the
Laniakea supercluster of galaxies, we place robust limits on the differences of the PPN parameter $\gamma$ values
for three cases, i.e., $\Delta\gamma<1.2\times10^{-10}$ for the polarized optical emission from GRB 120308A,
$\Delta\gamma<1.2\times10^{-10}$ for the polarized gamma-ray emission from GRB 100826A,
and $\Delta\gamma<2.2\times10^{-16}$ for the polarized radio emission from FRB 150807.

These are the first direct verifications of the WEP using multiband photons with different polarizations.
Moreover, the result from FRB 150807 provides the most stringent limit to date on the WEP. Compared with the
previous best limit from Crab pulsar photons ($\Delta\gamma \sim 10^{-15}$), which relied on using different
photon energies \cite{yang16}, our result represents an improvement of one order of magnitude.
If in the future the GRB polarimetric data are significantly enlarged by the gamma-ray polarimeter POLAR
on board the Chinese space laboratory Tiangong-II, and more FRBs with polarization information and redshift
measurements are detected by the Five Hundred Meter Aperture Spherical radio Telescope and the Square Kilometer
Array, much more stringent constraints on the WEP can be expected.

\acknowledgments
We are grateful to the anonymous referees for insightful comments.
This work is partially supported by the National Basic Research Program (``973'' Program)
of China (Grant No. 2014CB845800), the National Natural Science Foundation
of China (Grants No. 11433009, No. 11673068, and No. 11603076), the Youth Innovation
Promotion Association (2011231 and 2017366), the Key Research Program of Frontier Sciences (QYZDB-SSW-SYS005),
the Strategic Priority Research Program ``Multi-waveband gravitational wave Universe''
(Grant No. XDB23000000) of the Chinese Academy of Sciences, the Natural Science Foundation
of Jiangsu Province (Grant No. BK20161096), the Guangxi Key Laboratory for Relativistic Astrophysics,
and NASA NNX 13AH50G.

\bibliographystyle{apsrev4-1}

\begin{thebibliography}{0}%
\makeatletter
\providecommand \@ifxundefined [1]{%
 \@ifx{#1\undefined}
}%
\providecommand \@ifnum [1]{%
 \ifnum #1\expandafter \@firstoftwo
 \else \expandafter \@secondoftwo
 \fi
}%
\providecommand \@ifx [1]{%
 \ifx #1\expandafter \@firstoftwo
 \else \expandafter \@secondoftwo
 \fi
}%
\providecommand \natexlab [1]{#1}%
\providecommand \enquote  [1]{``#1''}%
\providecommand \bibnamefont  [1]{#1}%
\providecommand \bibfnamefont [1]{#1}%
\providecommand \citenamefont [1]{#1}%
\providecommand \href@noop [0]{\@secondoftwo}%
\providecommand \href [0]{\begingroup \@sanitize@url \@href}%
\providecommand \@href[1]{\@@startlink{#1}\@@href}%
\providecommand \@@href[1]{\endgroup#1\@@endlink}%
\providecommand \@sanitize@url [0]{\catcode `\\12\catcode `\$12\catcode
  `\&12\catcode `\#12\catcode `\^12\catcode `\_12\catcode `\%12\relax}%
\providecommand \@@startlink[1]{}%
\providecommand \@@endlink[0]{}%
\providecommand \url  [0]{\begingroup\@sanitize@url \@url }%
\providecommand \@url [1]{\endgroup\@href {#1}{\urlprefix }}%
\providecommand \urlprefix  [0]{URL }%
\providecommand \Eprint [0]{\href }%
\providecommand \doibase [0]{http://dx.doi.org/}%
\providecommand \selectlanguage [0]{\@gobble}%
\providecommand \bibinfo  [0]{\@secondoftwo}%
\providecommand \bibfield  [0]{\@secondoftwo}%
\providecommand \translation [1]{[#1]}%
\providecommand \BibitemOpen [0]{}%
\providecommand \bibitemStop [0]{}%
\providecommand \bibitemNoStop [0]{.\EOS\space}%
\providecommand \EOS [0]{\spacefactor3000\relax}%
\providecommand \BibitemShut  [1]{\csname bibitem#1\endcsname}%
\let\auto@bib@innerbib\@empty
\end{thebibliography}%


\begin{thebibliography}{99}

\bibitem{wil06} C.~M.~Will, Living Rev. Relativity {\bf 9}, 3 (2006);
{\bf 17}, 4 (2014).

\bibitem{sha64} I. I. Shapiro, Phys. Rev. Lett. {\bf 13}, 789 (1964).

\bibitem{Krauss88} L. M. Krauss and S. Tremaine, Phys. Rev. Lett. {\bf 60}, 176 (1988).

\bibitem{Longo88} M. J. Longo, Phys. Rev. Lett. {\bf 60}, 173 (1988).

\bibitem{gao15} H. Gao, X.-F. Wu, and P. M\'{e}sz\'{a}ros, Astrophys. J. {\bf 810}, 121 (2015).

\bibitem{wei2016} J.-J. Wei, X.-F. Wu, H. Gao, and P. M\'{e}sz\'{a}ros, J. Cosmol. Astropart. Phys. 08 (2016) 031.

\bibitem{Sang2016} Y. Sang, H.-N. Lin, and Z. Chang, Mon. Not. R. Astron. Soc. {\bf 460}, 2282 (2016).

\bibitem{wei15} J.-J. Wei, H. Gao, X.-F. Wu, and P. M\'{e}sz\'{a}ros, Phys. Rev. Lett. {\bf 115}, 261101 (2015).

\bibitem{tin16} S. J. Tingay and D. L. Kaplan, Astrophys. J. Lett. {\bf 820}, L31 (2016).

\bibitem{wei16} J.-J. Wei, J.-S. Wang, H. Gao, and X.-F. Wu, Astrophys. J. Lett. {\bf 818}, L2 (2016).

\bibitem{Wang2016} Z.-Y. Wang, R.-Y. Liu, and X.-Y. Wang, Phys. Rev. Lett. {\bf 116}, 151101 (2016).

\bibitem{yang16} Y.-P. Yang and B. Zhang, Phys. Rev. D {\bf 94}, 101501 (2016).

\bibitem{zhanggong16} Y.~Zhang and B.~Gong, Astrophys. J. {\bf 837}, 134 (2017);
S. Desai and E. O. Kahya, arXiv:1612.02532.

\bibitem{wu2016}
X.-F. Wu, H. Gao, J.-J. Wei, P. M\'{e}sz\'{a}ros, B. Zhang, Z.- G. Dai, S.-N. Zhang, and Z.-H. Zhu, Phys. Rev. D {\bf 94}, 024061 (2016).

\bibitem{Kahya2016}
E. O. Kahya and S. Desai, Phys. Lett. B {\bf 756}, 265 (2016);
X. Li, Y.-M. Hu, Y.-Z. Fan, and D.-M. Wei, Astrophys. J. {\bf 827}, 75 (2016).

\bibitem{nus16} A. Nusser, Astrophys. J. Lett. {\bf 821}, L2 (2016).

\bibitem{Zhang2016} S.-N. Zhang, arXiv:1601.04558;
Z.-X. Luo, B. Zhang, J.-J. Wei, and X.-F. Wu, J. High Energy Astrophys. {\bf 9-10}, 35 (2016).

\bibitem{Jackson1978} J.~D. Jackson, \emph{Classical Electrodynamics} (Wiley Eastern, New Delhi, 1978).

\bibitem{Roych2008} C. Roychoudhuri, A.~F. Kracklauer, and K. Creath, \emph{The Nature of Light. What is a Photon?}
(CRC Press, Boca Raton, FL, 2008).

\bibitem{Hofer1998} W.~A. Hofer, Physica A Statistical Mechanics and its Applications {\bf 256}, 178 (1998);
K. Muralidhar, Prog. Phys. {\bf 12}, 291 (2016).

\bibitem{di Serego Alighieri2015} A.~Lue, L.~Wang, and M.~Kamionkowski, Phys. Rev. Lett. {\bf 83}, 1506 (1999);
S. di Serego Alighieri, Int. J. Mod. Phys. D {\bf 24}, 1530016 (2015);
 W.-T. Ni, Int. J. Mod. Phys. Conf. Ser. {\bf 40}, 1660010 (2016).

\bibitem{footnote} Since our Milky Way galaxy locates at the corner of the Laniakea supercluster (see Fig.~2 of Ref. \cite{Tully2014}),
the gravitational potential of the light path from a GRB/FRB to us can be approximated as
a point mass potential for which all of the mass of Laniakea is assumed at the center of the mass.
Of course, this statement should be valid only if the light path is not close to the center of Laniakea.
Note that none of astrophysical transients we used in this manuscript is close to the center of Laniakea.

\bibitem{Tully2014} R.~B. Tully, H.~Courtois, Y.~Hoffman, and D.~Pomar{\`e}de, Nature (London) {\bf 513}, 71 (2014).

\bibitem{Lynden-Bell1988} D. Lynden-Bell {\it et al.}, Astrophys. J. {\bf 326}, 19 (1988).

\bibitem{Greiner2003} J.~Greiner {\it et al.}, Nature (London) {\bf 426}, 157 (2003).

\bibitem{Mundell2007} C. G.~Mundell {\it et al.}, Science {\bf 315}, 1822 (2007).

\bibitem{Steele2009} I. A.~Steele, C. G.~Mundell, R. J.~Smith, S.~Kobayashi, and C.~Guidorzi, Nature (London) {\bf 462}, 767 (2009).

\bibitem{Mundell2013} C. G.~Mundell {\it et al.}, Nature (London) {\bf 504}, 119 (2013).

\bibitem{Coburn2003} W.~Coburn and S. E.~ Boggs, Nature (London) {\bf 423}, 415 (2003).

\bibitem{Rutledge2004} R. E.~Rutledge and D. B.~Fox, Mon. Not. R. Astron. Soc. {\bf 350}, 1288 (2004);
C.~Wigger, W.~Hajdas, K.~Arzner, M.~G$\rm \ddot{u}$del, and A.~Zehnder, Astrophys. J. {\bf 613}, 1088 (2004).

\bibitem{Kalemci2007} E.~Kalemci, S. E.~Boggs, C.~Kouveliotou, M.~Finger, and M. G.~Baring, Astrophys. J. Suppl. {\bf 169}, 75 (2007).

\bibitem{McGlynn2007} S.~McGlynn {\it et al.}, Astron. Astrophys. {\bf 466}, 895 (2007).

\bibitem{Yonetoku2011} D.~Yonetoku {\it et al.}, Astrophys. J. Lett. {\bf 743}, L30 (2011).

\bibitem{Yonetoku2012} D.~Yonetoku {\it et al.}, Astrophys. J. Lett. {\bf 758}, L1 (2012).

\bibitem{Baumgartner2012} W. H. Baumgartner, D. N. Burrows, M. M. Chester, V. D'Elia, A. Y. Lien, C. B. Markwardt, K. L. Page, D. M. Palmer, and M. H. Siegel, GRB Coordinates Network {\bf 13017}, 1 (2012).

\bibitem{Deng2014} W.~Deng and B.~Zhang,  Astrophys. J. Lett. {\bf 783}, L35 (2014).

\bibitem{Lorimer2007} D.~R. Lorimer, M. Bailes, M.~A. McLaughlin, D.~J. Narkevic, and F. Crawford, Science {\bf 318}, 777 (2007);
E.~F. Keane, B.~W. Stappers, M. Kramer, and A.~G. Lyne, Mon. Not. R. Astron. Soc. {\bf 425}, L71 (2012);
D. Thornton \emph{et al.}, Science, {\bf 341}, 53 (2013);
S.~Burke-Spolaor and K.~W.~Bannister, Astrophys. J. {\bf 792}, 19 (2014);
L.~G. Spitler \emph{et al.}, Astrophys. J. {\bf 790}, 101 (2014);
V.~Ravi, R.~M.~Shannon, and A.~Jameson, Astrophys. J. Lett. {\bf 799}, L5 (2015);
D.~J. Champion {\it et al.}, Mon. Not. R. Astron. Soc. {\bf 460}, L30 (2016).

\bibitem{Keane2016} E.~F. Keane {\it et al.}, Nature (London) {\bf 530}, 453 (2016).

\bibitem{Williams2016} P. K. G. Williams and E. Berger, Astrophys. J. Lett. {\bf 821}, L22 (2016);
H. K. Vedantham {\it et al.}, Astrophys. J.  Lett. {\bf 824}, L9  (2016).

\bibitem{Chatterjee2017} S. Chatterjee {\it et al.}, Nature (London) {\bf 541}, 58 (2017).

\bibitem{Tendulkar2017} S. P. Tendulkar {\it et al.}, Astrophys. J. Lett. {\bf 834}, L7 (2017).

\bibitem{Petroff2015} E.~Petroff {\it et al.}, Mon. Not. R. Astron. Soc. {\bf 447}, 246 (2015).

\bibitem{Masui2015} K. Masui {\it et al.}, Nature (London) {\bf 528}, 523 (2015).

\bibitem{Ravi2016} V.~Ravi {\it et al.}, Science {\bf 354}, 1249 (2016).

\bibitem{web} http://www.astronomy.swin.edu.au/pulsar/frbcat/.

\bibitem{Cordes2016} J. M. Cordes, R. S. Wharton, L. G. Spitler, S. Chatterjee, and I. Wasserman, arXiv: 1605.05890.



\end{thebibliography}

\end{document}